\documentclass[5p,authoryear]{elsarticle}

\usepackage{amssymb}
\usepackage{mathtools}
\usepackage{latexsym}

\usepackage{url}

\usepackage{hyperref}
\usepackage[capitalise]{cleveref}
\usepackage{textcomp}
\usepackage{tabularx}
\usepackage{microtype}
\usepackage{siunitx}
\DeclareSIUnit\pixel{px}

\newcommand{\OO}{\mathcal{O}}
\providecommand{\CVLengthParam}{\nu}
\newcommand{\Per}{\operatorname{Per}}
\newcommand{\dx}{\operatorname{d\mathit{x}}}

\journal{Engineering Applications of Artificial Intelligence}

\begin{document}

\begin{frontmatter}

\title{Automation of Hemocompatibility Analysis Using Image Segmentation and a Random Forest}

\author[1]{Johanna C. Clauser\texorpdfstring{\corref{cor1}}{}}
\cortext[cor1]{Corresponding author: 
    mail: clauser@ame.rwth-aachen.de
    phone: +49-241-8089356; 
    fax: +49-241-8082144}
\author[1]{Judith Maas}
\author[1,5]{Jutta Arens}
\author[1]{Thomas Schmitz-Rode}
\author[1,2]{Ulrich Steinseifer}
\author[3,4]{Benjamin Berkels}

\address[1]{Department of Cardiovascular Engineering, Institute of Applied Medical Engineering, Medical Faculty,
RWTH Aachen University, Pauwelsstr. 20, 52074 Aachen, Germany}
\address[2]{Department of Mechanical and Aerospace Engineering, Faculty of Engineering, 
Monash Institute of Medical Engineering, Monash University, 17 College Walk, Clayton Victoria 3800, Australia}
\address[3]{AICES Graduate School, RWTH Aachen University, Schinkelstr. 2, 52062 Aachen, Germany}
\address[4]{Institute for Geometry and Practical Mathematics, RWTH Aachen University, Templergraben 55, 52056 Aachen, Germany}
\address[5]{University of Twente Faculty of Engineering Technology, Chair of Organ Support Technologies, 
 Department of Biomechanical Engineering, Drienerlolaan 5, 7522 NB Enschede, Overijssel, NL}

\begin{abstract}
The hemocompatibility of blood-contacting medical devices remains one of the major challenges
in biomedical engineering and makes research in the field of new and improved materials inevitable.
However, current in-vitro test and analysis methods are still lacking standardization and comparability,
which impedes advances in material design.
For example, the optical platelet analysis of material in-vitro hemocompatibility tests is carried out 
manually or semi-manually by each research group individually.

As a step towards standardization, this paper proposes an automation approach for the optical platelet count and analysis.
To this end, fluorescence images are segmented using Zach's convexification of the multiphase-phase piecewise constant Mumford--Shah model. 
The resulting connected components of the non-background segments then need to be classified as platelet or no platelet.
Therefore, a supervised random forest is applied to feature vectors derived from the components using features like area, perimeter and circularity. 
With an overall high accuracy and low error rates, the random forest achieves reliable results. 
This is supported by high areas under the receiver-operator and the prediction-recall curve, respectively.

We developed a new method for a fast, user-independent and reproducible analysis of material hemocompatibility tests,
which is therefore a unique and powerful tool for advances in biomaterial research.
\end{abstract}


\end{frontmatter}

\section{Introduction}
\label{sec1}
The major challenge in the field of blood contacting medical devices like e.g.\ artificial heart valves or 
circulatory support systems remains the biocompatibility of these materials.
With regard to blood contact, the subtopic of hemocompatibility is of major interest \citep{Ratner2007}.
When a foreign material gets into contact with blood,
proteins are adsorbed on the material surface in the first instance.
In the following, the physiological coagulation system is triggered and activated,
which finally leads to platelet adhesion and thrombus formation on the surfaces \citep{Jaffer2015}.
Possible consequences of such thrombi are failure of the medical device or thromboembolism,
which are both highly life-threatening for the patients.  
An anticoagulation therapy can reduce the risk of thrombi, however, it is accompanied by possible bleeding complications \citep{Ratner2013}.
Due to these limitations in the use of artificial blood contacting materials, research and improvements
in the field of material hemocompatibility is of high importance.

The evaluation of a material's hemocompatibility is carried out at first in in-vitro blood tests.
A general framework for such tests is defined in the standard DIN ISO 10993-4.
However, it consists more of recommendations rather than of clear instructions or regulations \citep{Braune2015, ISO_10993-4_2017}.
As an example, the ISO suggests an optical evaluation of adherent platelets after material in-vitro testing by means of different microscopy techniques
but does not provide any general instructions how the degree of platelet adhesion or activation should be evaluated.
Thus, there is a huge variability of applied test and analysis methods that make comparisons between different studies nearly impossible \citep{Braune2013, Sefton2001}.
In the case of microscopy image analysis, a manual or semi-manual count of adherent platelets is a commonly used tool \citep{Lutter2015, Pham2016, Zhou2008}. 
Beside the lack of transferability of such user-dependent analyses, a (semi-)manual platelet count can be very time consuming and is prone to errors or personal opinions.
An automatized platelet analysis would overcome these limitations and allow for a high-quality and reproducible microscopic hemocompatibility assessment.

This study presents a new approach for an automatized analysis of fluorescence images for in-vitro hemocompatibility evaluation 
based on a piecewise-constant multiphase segmentation algorithm and a supervised random forest.

\section{Methods}
\label{sec2}
The first part of this chapter contains information about the test setup and the experimental procedure of the in-vitro
hemocompatibility study that was carried out for creating a fluorescence image data base.
In the second part, mathematical details of the segmentation algorithm are presented.
The third part comprises information regarding the supervised random forest
and in the last part of this chapter, a multi-person prediction trial is described.

\subsection{Hemocompatibility Experiments}
\label{sec2.1}
The hemocompatibility test series was carried out with three commercial medical grade materials as well as 
two different in-house produced foils of medical grade polyurethane (see \cref{tab1:Materials}). 
Materials were chosen due to comparability to previously conducted studies with these materials \citep{Braune2017, Clauser2014, Nadzeyka2017}.
The experimental procedure of the static in-vitro experiments is described in detail in \citep{Braune2017}.
Briefly, duplicates of material samples were incubated for \SI{1}{\hour} each with \SI{1}{\milli\litre} of platelet rich plasma (PRP) 
obtained from fresh human blood (Ethical Committee reference number EK 348/16 and EK 033/18).
Afterwards, PRP was removed and material samples were consecutively rinsed with phosphate buffered saline,
fixed with \SI{2}{\percent} glutaraldehyde (Roth, Germany), rinsed with buffer solution again and 
finally mounted on microscopy slides (Mowiol, Roth, Germany).
Experiments were repeated ten times with different donors.

\begin{table*}[!t]
  \caption{\label{tab1:Materials}Materials for hemocompatibility testing}
  \centering
  \begin{tabular}{l|l|l}
  \hline
  \textbf{ID} & \textbf{Material} & \textbf{Manufacturer} \\
  \hline
  M1 & PDMS: Poly(dimethylsiloxane) & Bess Medizintechnik GmbH, Germany\\
  M2 & PTFE: Poly(tetrafluoro ethylene) & Bess Medizintechnik GmbH, Germany\\
  M3 & PET: Poly(ethylene terephthalate) & ThermoFisher Scientific, USA\\
  M4 & PCU\textsubscript{extr}: Poly(carbonate urethane) -- extrusion & Lubrizol, USA (raw material)\\
  M5 & PCU\textsubscript{cast}: Poly(carbonate urethane) -- mold casting & Lubrizol, USA (raw material)\\
  \hline
\end{tabular}
\end{table*}

Microscopy images were acquired with a fluorescence microscope
(Axio Observer Z1, Carl Zeiss GmbH, Germany) and a 40-fold magnification.
The images were recorded as 5x5~tile images with a maximum of 45~tile images distributed across the material sample.
For each 5x5~tile, nine focus points were adjusted manually prior to image recording.
Regions of contamination (e.g.\ mounting medium, air bubbles, glutaraldehyde residues) were excluded from microscopy.
Due to a high autofluorescence of PET, light intensity was reduced for this material to \SI{58.32}{\percent}
compared to the other materials recorded with \SI{100}{\percent} light intensity.
Following microscopy, the region of interest (ROI) representing the PRP-incubated area was marked 
on a compilation of all images of one material samples using the software Zen Blue (Carl Zeiss GmbH, Germany).
Any image outside the ROI was excluded from the following analysis steps.
Additionally, all remaining images underwent a visual inspection,
discarding blurred or contaminated (e.g.\ hair, protein residues) images.
In the following, a subset of the remaining images was used for the development of the new analysis method (ground truth), 
which was afterwards applied to the whole image data set for analyzing the hemocompatibility study.

\subsection{Image Segmentation}
\label{sec2.2}
\begin{figure}[!b]
\centering
\begin{tabular}{cc}
\includegraphics[width=.45\linewidth]{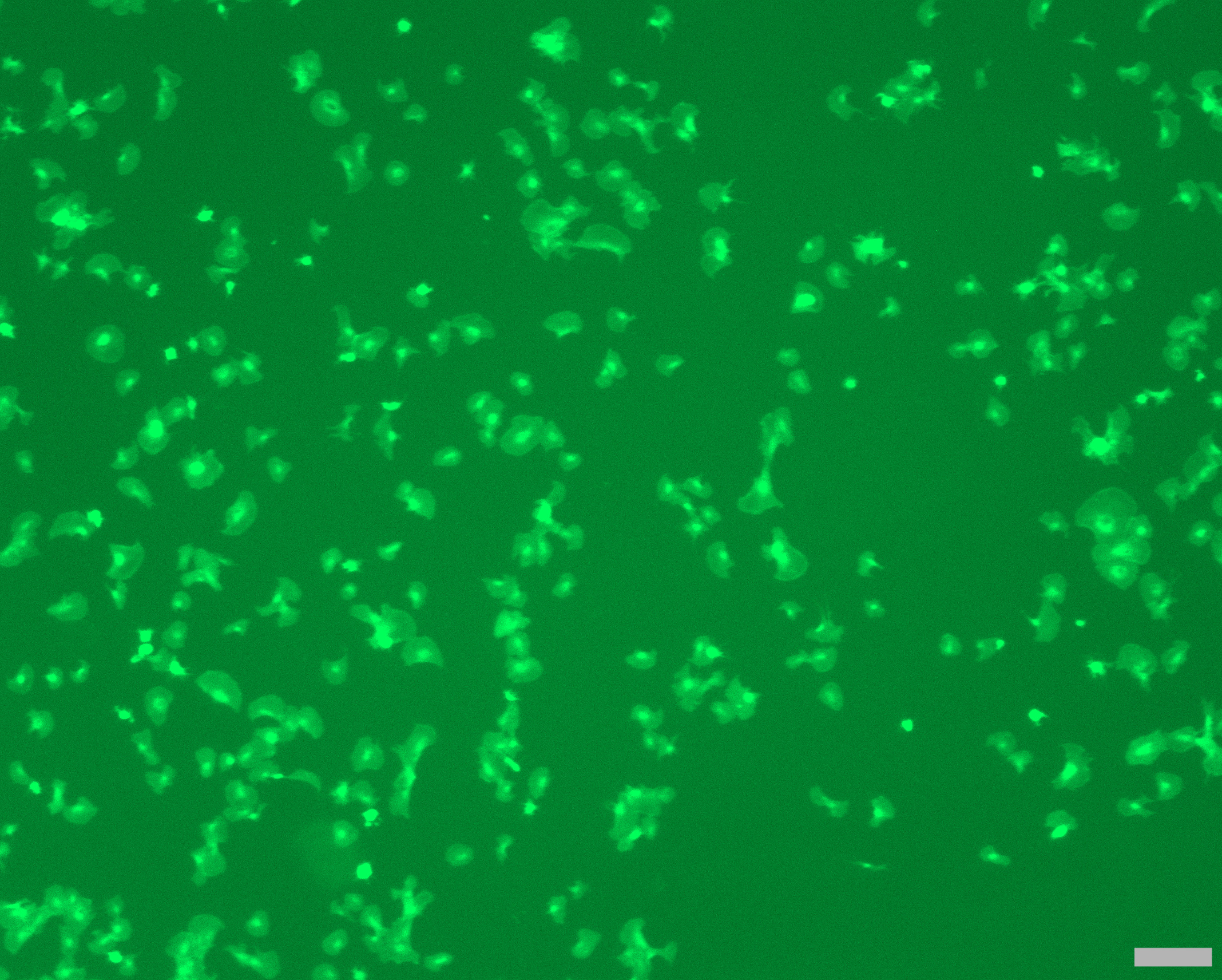}&
\includegraphics[width=.45\linewidth]{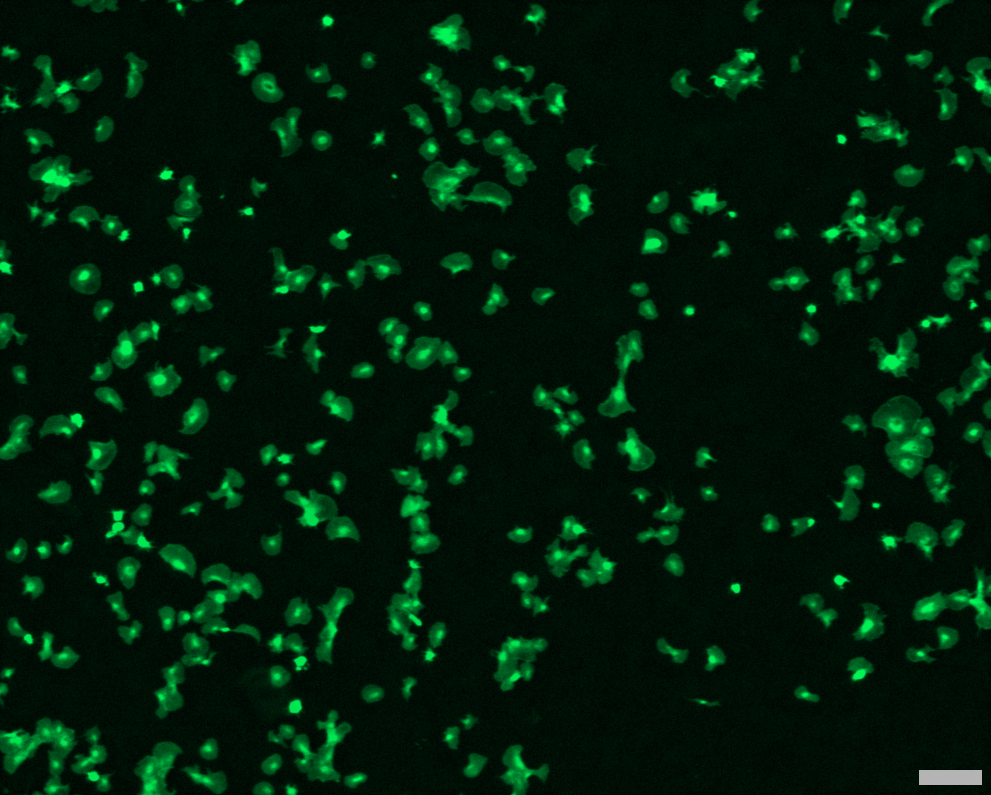}
\end{tabular}
\caption{A fluorescence image before (left) and after (right) background correction with the rolling ball algorithm; scale bar $\hat{=}$ \SI{20}{\micro\metre}}
\label{fig:RollingBall}
\end{figure}

The first step in the platelet analysis of an image is to detect all objects shown in that image as a basis to find potential platelets.
To simplify this detection, we first removed artifacts and the uneven background illumination from the images using ImageJ (National Institutes of Health, USA).
For this, the rolling ball algorithm was applied with a radius of $r=\SI{100}{\pixel}$,
which achieved the best results according to visual verification.
Additionally, images were downscaled by the factor 2.777 (final resolution of $991\times\SI{795}{\pixel}$)
to accelerate the following processing steps.
\cref{fig:RollingBall} shows an example image before and after applying the rolling ball algorithm.

As one can see in the background corrected image in \cref{fig:RollingBall}, finding potential platelets essentially means separating foreground from background in the image and splitting the foreground into its connected components. This separation is a special image segmentation problem.
In general, image segmentation refers to the task of partitioning an image into segments that describe image structures, e.g.\ foreground and background. More mathematically, it means decomposing an image into disjoint regions that are homogeneous in a suitable sense, e.g.\ that have a homogeneous color. As one of the fundamental image processing problems, it has been extensively studied in the literature in many contexts and for many applications. For instance, a survey on trends in color image segmentation can be found in \citep{VaSa12}.
Here, we use the the famous Mumford--Shah model to handle the segmentation \citep{MuSh89}.

\begin{figure}[!b]
  \centering
  \begin{tabular}{cc}
  \includegraphics[width=.45\linewidth]{Figures/Fig_ExampleBackground}&
  \includegraphics[width=.45\linewidth]{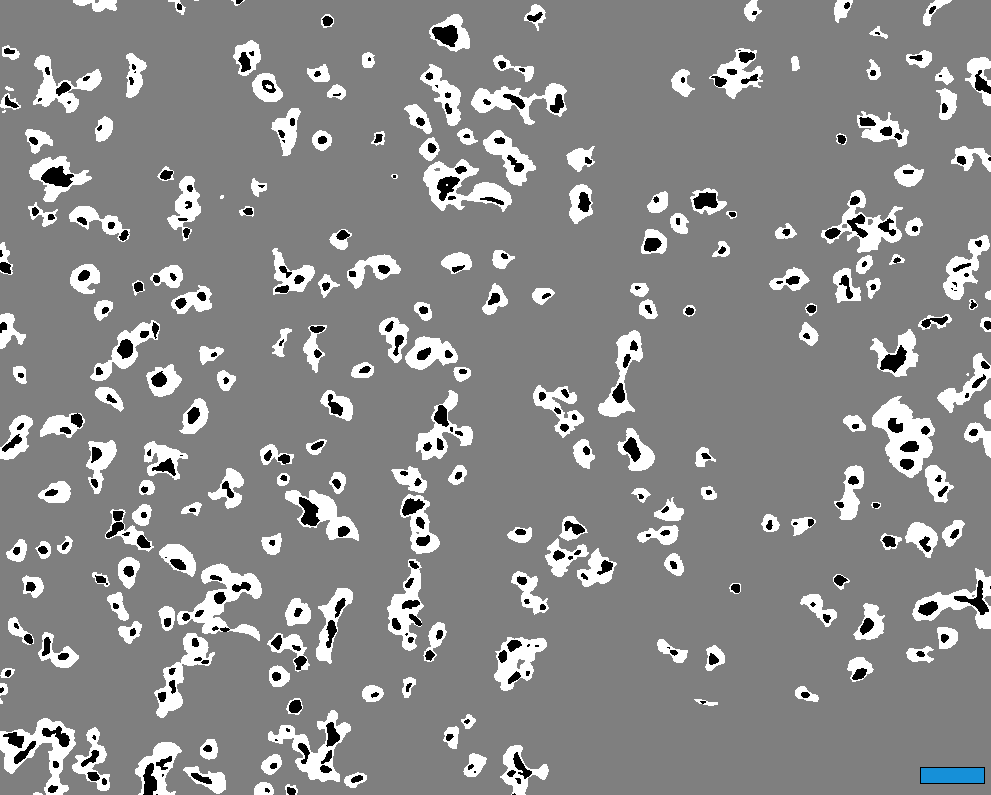}
  \end{tabular}
  \caption{A fluorescence image before (left) and after (right) image segmentation; scale bar $\hat{=}$ \SI{20}{\micro\metre}}
  \label{fig:Segmentation}
  \end{figure}

Mathematically, an image is a mapping from an image domain $\Omega\subset\mathbb{R}^2$ (here just a rectangle) to a range of values, e.g. real valued vectors of length three $\mathbb{R}^3$ for color images. Let $g$ be an image, i.e. a mapping as just described. Now, the famous piecewise-constant Mumford--Shah model approximates $g$ with an image that only has a fixed number $L$ of different color values. This is done by minimizing
\begin{equation*}
  E[(\OO_l)_{l=1}^L,(c_l)_{l=1}^L] = \sum_{l=1}^L \left\lbrace \int_{\OO_l} (g(x)-c_l)^2\,\dx + \CVLengthParam \Per(\OO_l) \right\rbrace,
\end{equation*}
where $\OO_1,\ldots,\OO_L$ is a partition of $\Omega$, e.g. the sets are pairwise disjoint and the union of their closures is equal to the closure of $\Omega$, i.e. $\bigcup_{j=1}^n \bar{\OO}_j = \bar\Omega$, and, for $l=1,\ldots,L$, the vector $c_l\in\mathbb{R}^3$ is the mean color of $g$ in the segment $\OO_l$. $\Per$ denotes the perimeter, i.e. the length of the boundary.
Essentially, this means finding segments $\OO_l$ and corresponding color values $c_l$, such that the color values of $g$ in a segment $\OO_l$ are close to the color $c_l$, while trying to keep the boundary length of the sets $\OO_l$ short. This latter is necessary to avoid very irregular segments and very small segments, which could be caused by noise in $g$.
The corresponding minimization problem is non-convex and thus difficult to solve. 
We estimate the color values $c_l$ by $k$-means clustering of the color values of $g$ with $L$ clusters \citep{Ma67}. 
As in reference \citep{MeBe16}, we estimate the sets $\OO_l$ using the convexification of the Mumford--Shah segmentation approach from \citep{ZaGaFr08}. 
The resulting convex optimization problem is solved with the primal-dual hybrid gradient method~\citep{ChPo11}.
\cref{fig:Segmentation} shows the result of the segmentation on the background corrected image from \cref{fig:RollingBall} with $L=3$, i.e. when approximating the input image with three different color values.
One of the color values is used to approximate the black background, the other two are used to approximate the potential platelets with two different shades of green.
This leads to a clear separation of the potential platelets from the background combined with a further separation of the the potential platelets into brighter and darker areas, which are shown, respective, as black and white pixels in the right part of \cref{fig:Segmentation}).
Using connected components labeling on the non-background pixels \citep{Sh96a}, we extract the connected components of the result. 
Each of those is a potential platelet and needs to be classified as platelet or no platelet. 
For this, we derive a feature vector that contains the necessary information for the classification.
The feature vector is computed for each component and consists of the total number of pixels (area) of the component, the number of brighter and darker image pixels in the component, the perimeter of the component  and its circularity. 
The latter is defined as $\frac{P^2}{4\pi A}$, where $P$ is the perimeter and $A$ the area of the component.
Additionally, the number of brighter and darker connected regions in each component and the ratio of bright to dark regions is computed for every component. 
Theses feature vectors are used as input for the classification of each component.

\subsection{Random Forest and Prediction Statistics}
\label{sec2.3}
\begin{figure}[!t]
\centering
\includegraphics[width=.40\linewidth]{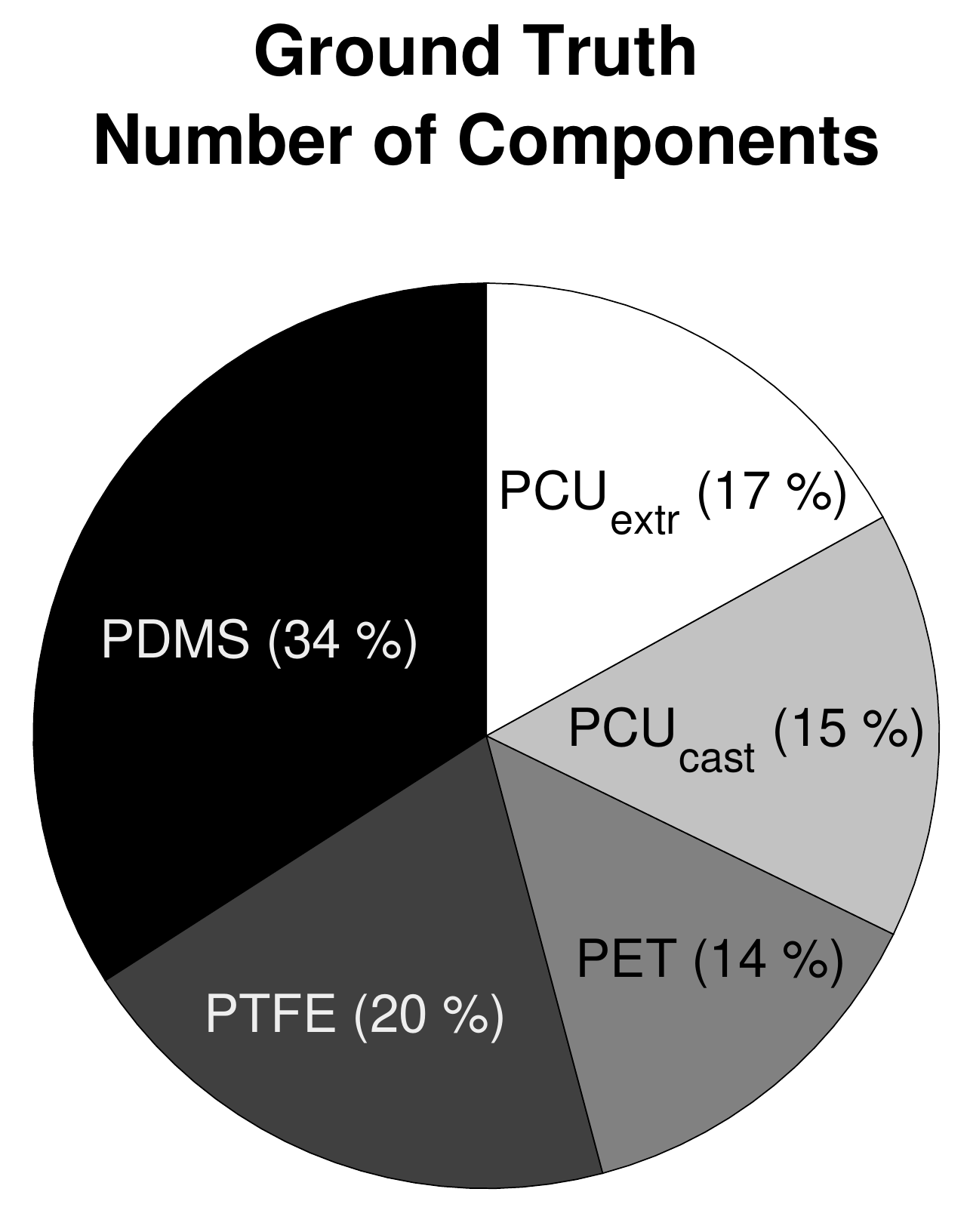} \quad\includegraphics[width=.40\linewidth]{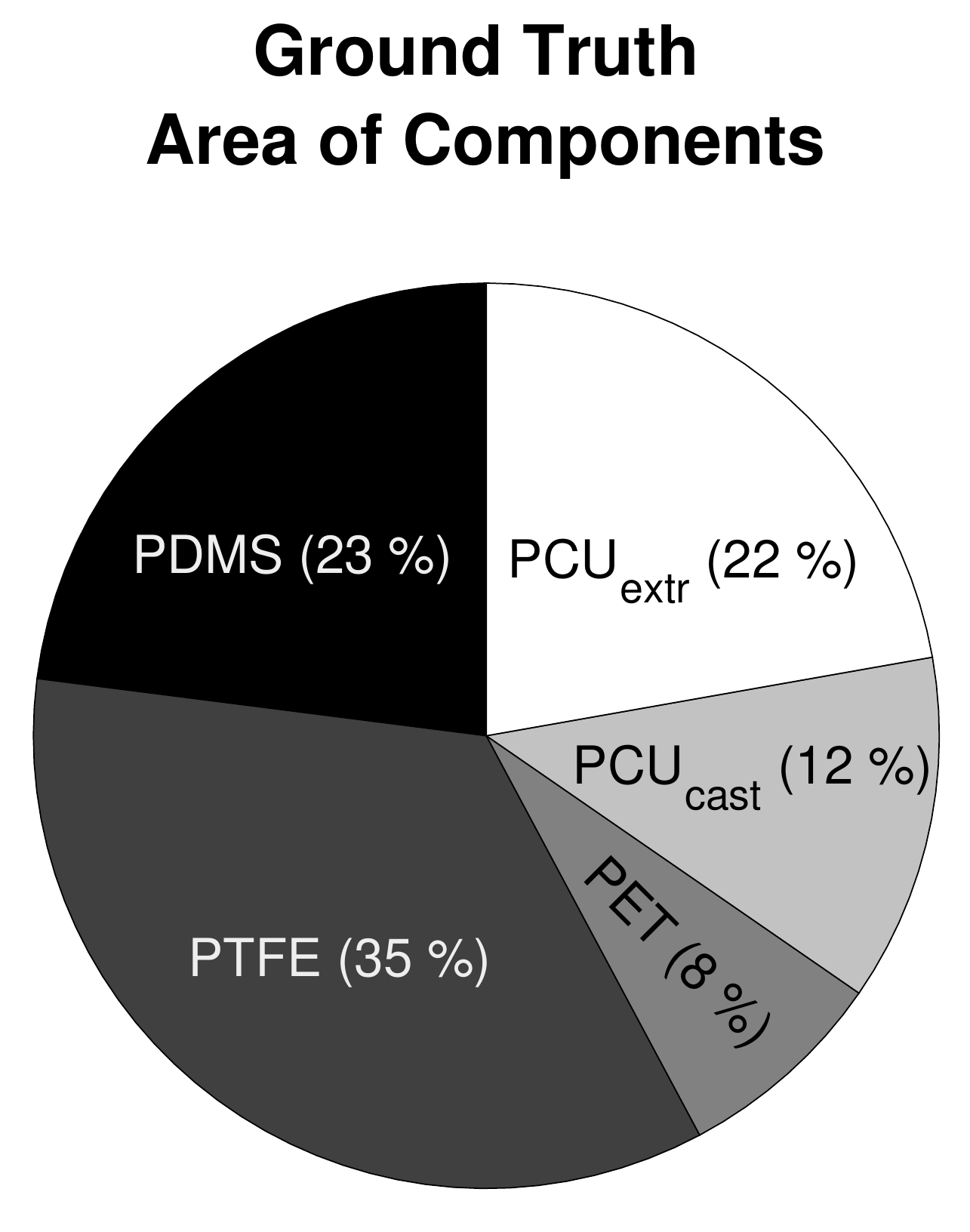}
\caption{Ground truth data distribution for the number of components (left) and the area of components (right)}
\label{fig:AreaCompPerMat}
\end{figure}

For the final analysis of the number and the area of adherent platelets on the microscopy images,
a machine learning algorithm was set up.
We used the implementation for the random forest algorithm from software package KNIME (KNIME AG, Switzerland).
Training data for a supervised random forest (RF) model was obtained
by manual classification of 19 628 components with an overall area of \SI{7257484}{\pixel} (\SI{0.0997}{\micro\metre\squared\per\pixel}).
These included \SI{22}{\percent} 'no platelets', indicated as the positive class and accordingly
\SI{78}{\percent} 'platelets', the negative class.
As the platelet morphology is impacted by the material the platelet adheres to, 
components from each material group were characterized.
Note that aggregates formed by several platelets were considered as one component.
Thus, the resulting area of components differs from the distribution of the number of components.
\cref{fig:AreaCompPerMat} shows the distribution among the materials for both the number and the area of components.

The manually classified components were divided into training, validation and test data.
In each subset, the above-mentioned ratio of positive and negative components was maintained. 
\cref{tab2:ClassificationData} shows the percentage of each data group for the number and the area of components.

\begin{table}[!b]
  \caption{\label{tab2:ClassificationData}Learning data sets for the random forest}
  \centering
  \begin{tabular}{p{0.28\columnwidth}|p{0.28\columnwidth}|p{0.28\columnwidth}}
  \hline
  \textbf{Data Set} & \textbf{Number of \newline Components $($\%$)$} & \textbf{Area of \newline Components $($\%$)$} \\
  \hline
  Training Data & 71.5 & 74.4\\
  Validation Data & 17.9 & 18.5\\
  Test Data & 10.6 & 7.1\\
  \hline
\end{tabular}
\end{table}

The training data was used to set up a RF with specific parameters and the resulting RF was used to classify the validation data. 
Results were compared to the manual classification (ground truth) and prediction quality was evaluated by means of:
\begin{itemize}   
  \item accuracy (ACC)
\end{itemize}
\begin{equation}
  \label{eq:ACC}
  ACC=\frac{TP+TN}{TP+FP+TN+FN}
\end{equation}

\begin{itemize}
  \item Mathews correlation coefficient (MCC)
\end{itemize}
\begin{equation}  
  \label{eq:MCC}
  MCC=\frac{TP\cdot TN-FP\cdot FN}{\sqrt{(TP+FP)(TP+FN)(TN+FP)(TN+FN)}}
\end{equation}

\begin{itemize}
  \item false positive rate (FPR)
\end{itemize}
\begin{equation} 
  \label{eq:FPR} 
  FPR=\frac{FP}{TN+FP}={1-\text{specificity}}
\end{equation}

\begin{itemize}  
  \item false negative rate (FNR)
\end{itemize}
\begin{equation}
  \label{eq:FNR} 
  FNR=\frac{FN}{TP+FN}={1-\text{sensitivity}}
\end{equation}
\begin{tabularx}{\columnwidth}{XX}
  \textit{TP}: true positive; & \textit{TN}: true negative \\
  \textit{FP}: false positive; & \textit{FN}: false negative \\
\end{tabularx}

\bigskip
This procedure was repeated adjusting the RF parameters until the maximum of prediction quality was reached.
Adjustable parameters were split criterion, number of levels (tree depth), minimum node size
and number of nodes.
The following configuration achieved the best prediction outcome:
\begin{itemize}   
  \item split criterion:  information gain ratio (NI)
\end{itemize}
\begin{equation}
  NI=\frac{H(R)-\frac{|R_l|}{|R|}H(R_l)-\frac{|R_r|}{|R|}H(R_r)}{-\left(\frac{|R_l|}{|R|}\log_2\frac{|R_l|}{|R|}+\frac{|R_r|}{|R|}\log_2\frac{|R_r|}{|R|}\right)}
\end{equation}
  where $H$ denotes the Shannon entropy
\begin{equation}
  H(R)=-\sum_{k=1}^Kp_k\log_2p_k
\end{equation}
\begin{tabularx}{\columnwidth}{lX}
  $p_k$: probability of $k$\textsuperscript{th} class in $R$; & $R_l$: population in left child node \\
  $R$: total population; & $R_r$: population in right child node \\
\end{tabularx}

\begin{itemize}
  \item number of levels: 100
  \item minimum node size: unconstrained
  \item number of nodes: unconstrained
\end{itemize}

Additionally to validation data prediction, a 10-fold cross validation was done
to ensure the model to be independent from the training data set.
To this end, the union of training and validation data was randomly divided into 10 subsets of data, 
each with a positive/negative ratio of \SI{21.45}{\percent}/\SI{78.55}{\percent}.
A RF with the parameters described above was grown on 9~folds and validated against the ground truth on the remaining fold.
This procedure was repeated 10~times, each time with another remaining validation fold. 

Finally, the union of training and validation data was used to grow the final random forest.
To evaluate the performance of the final RF, the so-far untouched test data set was predicted by this RF and the corresponding prediction statistics were computed.
To further qualify the test data prediction, the Receiver-Operator-Characteristic (ROC) curve and the Precision-Recall (PR) curve, which are based on
recall \eqref{eq:recall}, fallout \eqref{eq:fallout} and precision \eqref{eq:precision}, were evaluated. 
The ROC curve represents recall vs fallout, whereas the PR curve depicts precision vs recall.
The latter thereby takes an over-representation of the negative class more into account \citep{Touw2013}.

\begin{equation}
  \label{eq:recall}
  \text{recall}=\text{sensitivity}=\frac{TP}{TP+FN}
\end{equation}
\begin{equation}
  \label{eq:fallout}
  \text{fallout}=FPR
\end{equation}
\begin{equation} 
  \label{eq:precision}
  \text{precision}=\frac{TP}{TP+FP}
\end{equation}

\subsection{Multi-Person Prediction}
\label{sec2.4}
Supervised prediction models are subject to the risk of being biased by the person who has provided the learning data.
The independence of the RF from the underlying data was verified in a multi-person prediction trial.
Five people who were not involved in data collection before did a manual classification of a smaller subset of the ground truth data,
including 1\,045 components with an area of \SI{260922}{\pixel}. 
The distribution between positive (\SI{78}{\percent}) and negative (\SI{22}{\percent}) components was kept similar to the ground truth ratio.
Prediction statistics were calculated for every pairwise combination of the seven available classifications,
namely the RF, the original classifying person ('P1') and the five additional persons ('P2' to 'P6').
Note that the combination RF-P1 thus represents a part of the earlier calculated statistics during setting up the RF.

\section{Results}
\label{sec3}

This section first presents the random forest statistics for the algorithm itself, followed by the multi-person statistics.
Finally, the newly developed, automated method is used to analyse the conducted in-vitro hemocompatibility experiment as a proof-of-concept.

\begin{table*}[!t]
  \caption{\label{tab3:10foldResults}10-fold cross validation for the number and the area of components}
  \centering
  \begin{tabular}{l|l|l|l|l|l|l}
  \hline
   & \textbf{P $($\%$)$} & \textbf{N $($\%$)$} & \textbf{ACC} & \textbf{MCC} & \textbf{FPR $($\%$)$} & \textbf{FNR $($\%$)$} \\
  \hline
  \textbf{\# Mean} & 21.45 & 78.55 & 0.96 & 0.87 & 1.97 & 12.82 \\
  \textbf{\# STD} & $\pm$\,0.02 & $\pm$\,0.02 & $\pm$\,0.01 & $\pm$\,0.01 & $\pm$\,0.45 & $\pm$\,1.55 \\
  \hline
  \textbf{A Mean} & 7.23 & 92.77 & 0.96 & 0.70 & 0.62 & 41.11 \\
  \textbf{A STD} & $\pm$\,0.93 & $\pm$\,0.93 & $\pm$\,0.01 & $\pm$\,0.07 & $\pm$\,0.38 & $\pm$\,9.23 \\  
  \hline
\end{tabular}
\end{table*}
\subsection{Random Forest Statistics}
\label{sec3.1}

The results of the 10-fold cross validation are shown in \cref{tab3:10foldResults} for both the number (\#) and the area (A) of components.
The confusion matrix values are not shown as mean values and standard deviations are not reasonable for these numbers.

The distribution of positive (P) and negative (N) components is extremely imbalanced for the area of components
as it was not specifically set.
Nevertheless, values for ACC and MCC are in a similar high range for the number and the area of components.
FPR is low for both cases with the higher value for the number of components with $1.97 \pm 0.45$.
FNR is considerably higher, especially for the area of components ($41.11 \pm 9.23$).
This means that about \SI{40}{\percent} of the component area of positive components (no platelets) is characterized as the negative class (platelet).
In general, all parameter show small standard deviations throughout the 10-fold cross validation.

\cref{tab4:TestDataPrediction} shows the prediction results of the test data by the final RF model for the number and the area of components.
ACC and MCC are in a high range above 0.9 and 0.7, respectively, with slightly lower MCC for the area of components compared to the number of components.
The FPR for the number of components is with \SI{5.13}{\percent} higher than for the area of components (\SI{1.13}{\percent}).
By contrast, FNR is lower for the number of components (\SI{11.83}{\percent}), nevertheless, 
FNR for the area of components (\SI{32.62}{\percent}) is lower compared to the 10-fold cross validation.
Note that the positive/negative component ratio is shifted again towards the negative class for the area of components.
\begin{table}[!b]
  \caption{\label{tab4:TestDataPrediction}Test Data Prediction}
  \centering
  \begin{tabular}{l|l|l}
  \hline
  & \textbf{Number} & \textbf{Area $($px$)$} \\
  \hline
  \textbf{TP} & 410 & 23\,652 \\
  \textbf{FP} & 83 & 5\,432 \\
  \textbf{TN} & 1\,535 & 473\,422 \\
  \textbf{FN} & 55 & 11\,451 \\
  \textbf{P/\%} & 22.32 & 6.83 \\
  \textbf{N/\%} & 77.68 & 93.17 \\
  \textbf{ACC} & 0.93 & 0.97 \\
  \textbf{MCC} & 0.81 & 0.72 \\
  \textbf{FPR/\%} & 5.13 & 1.13 \\
  \textbf{FNR/\%} & 11.83 & 32.62 \\
  \hline
  \end{tabular}
\end{table}

Distinguishing the test data statistics with regard to the five different materials (cf. Supplementary Material, Table S1 \& Table S2), the number of components shows only minor difference in prediction quality between all materials.
ACC varies from 0.90 (PDMS) to 0.98 (PCU\textsubscript{extr}) and MCC varies from 0.78 (PDMS) to 0.88 (PCU\textsubscript{extr}), which is in accordance with the total statistics.
PDMS reveals the highest FPR (\SI{5.39}{\percent}) and FNR (\SI{4.47}{\percent}).
For the area of components, ACCs are slightly higher and MCCs are slightly lower compared to the number of components, which is in line with the previous results.
FPRs are in general lower than for the number of components, whereas the FNR for PDMS (\SI{33.57}{\percent}) and PET (\SI{32.31}{\percent}) are in the high range of the total statistics.

To further quantify the prediction quality, the ROC curve and PR curve of the test data were calculated.
The area under the ROC curve is 0.98
and the area under the PR curve is 0.85
for the number of components.
For the area of components, results are similar with 0.94 and 0.77 for the area under the ROC and the PR curve, respectively.
The graphs are provided in the supplementary (Fig. S1).

Detailed data for the multi-person prediction is shown in the supplementary material (Tables S3-S6).
Comparing each classifier with each other shows only marginal differences and no tendencies.
All ACCs are above 0.90 for both the number and the area of components, which is in line with all previous results.
For some combinations, MCCs are slightly lower than the total RF statistics, however, the values are never lower than 0.62 (P5-RF).
FPRs do not exceed \SI{5}{\percent} for any combination, whereas FNRs range up to \SI{24.05}{\percent} (P1-P4) for the number of components and
\SI{49.46}{\percent} (P5-RF) for the area of components.

\subsection{Hemocompatibility Analysis}
\label{sec3.2}
\begin{figure}[!b]
  \centering
  \includegraphics[width=0.9\linewidth]{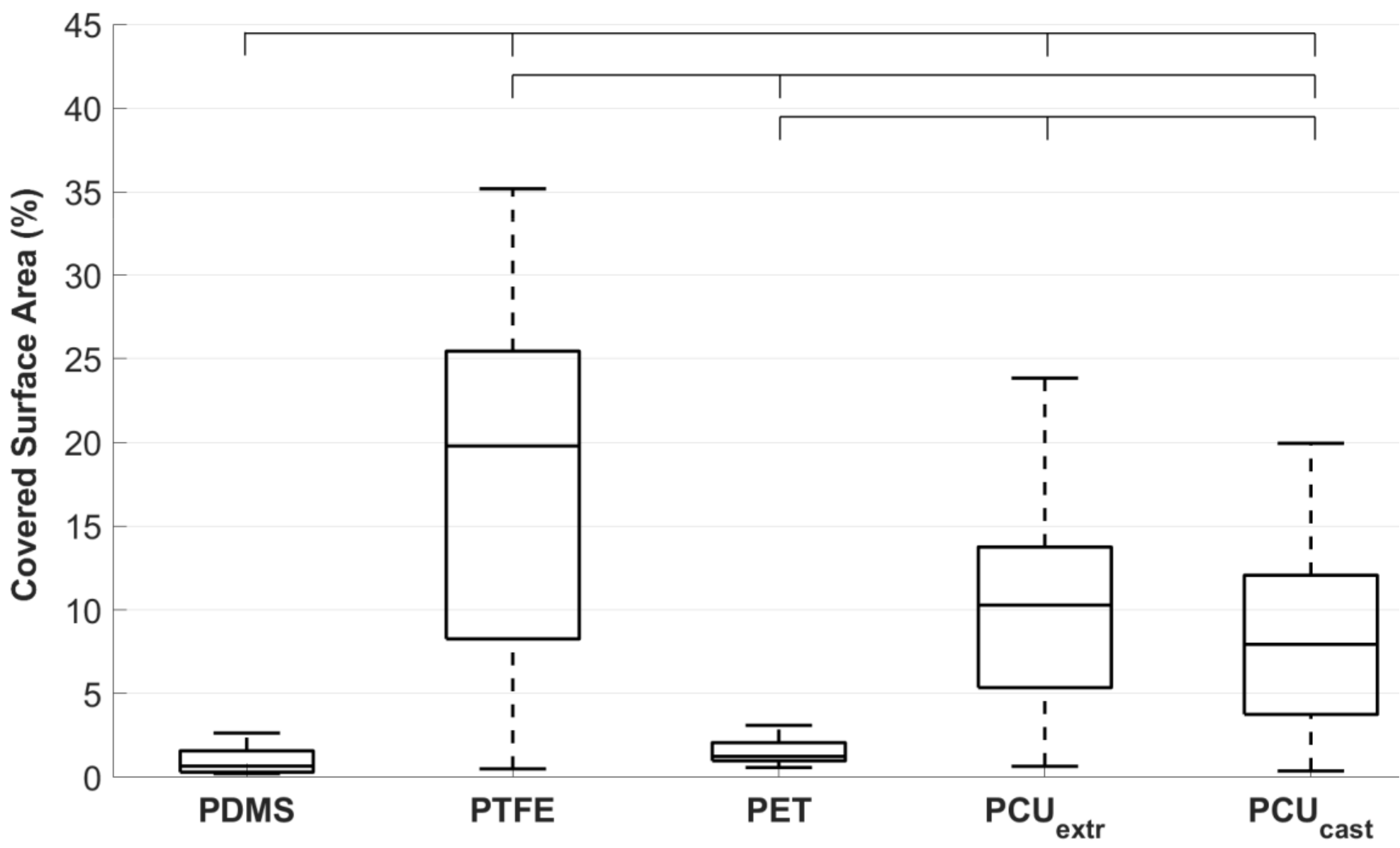}
  \caption{Covered surface area, crossbar for \textit{p}\,$<$\,0.05}
  \label{fig:CoveredArea}
  \end{figure}

The above developed and validated segmentation algorithm and the RF model were used to run a complete analysis of the hemocompatibility study.
\cref{fig:CoveredArea} shows the platelet covered surface area for the five different materials. 

Cross bars indicate significant differences ($p<0.05$), 
calculated by one-way analysis of variance for normally distributed data sets and Kruskal-Wallis test for non-normally distributed data, respectively (IBM SPSS Statistics~24, IBM, USA).
PTFE shows the most covered surface area and the most scattered values with $\SI{18.75}{\percent} \pm \SI{9.50}{\percent}$.
Both PCU materials are in the range of \SI{10}{\percent} covered surface area and the least covered surface areas are found on PDMS and PET, both in the range of \SI{1}{\percent}.
Except of the combinations PDMS-PET and PTFE-PCU\textsubscript{extr}, all differences between the materials are significant.

\section{Discussion}
\label{sec4}
A new analysis method for in-vitro hemocompatibility tests was developed, 
including the automation of image segmentation and platelet analysis using a RF algorithm.

The quality of the image segmentation was evaluated by visual inspection, 
by comparing the original and the segmented image of different microscope images from each material group.
The main criterion was the detailed depiction of platelet pseudopodia while maintaining the original platelet area.
Although this is a manual interaction step, it has to be done only once during the set up of the algorithm and is thus acceptable.
Continuous control of images throughout the whole study proved the segmentation to be precise for a large number of diverse images ($>$\,200) and thus a possible error based on the manual interaction to be very small. 

The RF for platelet characterization was set up in a supervised manner.
To avoid any biasing by manual image selection, images were chosen randomly from all five material groups.
The large number of classified components (nearly \num{20000}) ensures a good representation of all data by the ground truth.
As PDMS turned out to be the most critical material in terms of protein residues, more components were classified from this material group.
For the evaluation of a material's hemocompatibility, the area covered with platelets is a crucial parameter.
Thus, not only the number but also the area of components is evaluated by the RF.
The area of one 'component' can differ immensely depending on the type of component.
A single platelet covers an area of around \SI{10}{\micro\metre\squared} whereas a platelet aggregated consisting of numerous connected platelets could cover a whole microscope image.
Accordingly, positive-negative class ratios are shifted towards the negative class as the distribution was set for the number of components.
This is the reason why prediction values that take the here-in underrepresented positive class more into consideration 
are slightly lower for the area of components than for the number of components.
Nevertheless, all prediction statistics reveal ACC and MCC considerably above 0.5 and 0, respectively, 
which is the threshold indicating a well-founded prediction rather than a random choice.
This outcome is supported by the areas under the ROC and the PR curve, which are all close to 1.
RF statistics cannot be compared to similar studies as there are no reported approaches for platelet characterization by means of segmentation and machine learning methods.
Instead, studies using RF models in the field of biomedical imaging (e.g.\ CT, disease marker) are used for comparison.
With regard to ACC and MCC, our model shows prediction values comparable or even superior to other reported RFs \citep{Jiang2007, Gray2013a, Wu2009d, Li2012a, Liu2010b, Desir2012}.
The same outcome is observed for ROC and PR curve evaluation \citep{Liu2010b, Khalilia2011, Statnikov2008, Lempitsky2009}.
These results prove that our proposed RF model has an overall high prediction quality, which is in agreement with other biomedical applications.

Having a more detailed look at the error rates, considerably higher FNRs than FPRs are observed.
This means, a certain amount of 'no platelets' are characterized as platelets.
In the context of material thrombogenicity assessment, this represents a worst-case-scenario in terms of hemocompatibility rating.
Although the aim is the most perfect characterization as possible, however,
a slight overrating in terms of thrombogenicity can be considered as an additional level of safety.
The underestimation of a material's thrombogenicity can have a severe and even life-threatening impact on patients, thus, 
realistic predictions are crucial.
For FPRs, values are in the range of \SI{5}{\percent} or lower, which represents a very small error rate for platelets characterized as no platelets.
Interestingly, other studies that reported the sensitivity as the counterpart of FPR and the specificity as the counterpart of FNR 
present a similar phenomenon \citep{Gray2013a, Li2012a, Liu2010b, Wu2009d}.
Possibly, high FNRs can be attributed to the field of biomedical imaging, where the over-represented negative class contains characteristic components,
whereas the positive class can appear in a variety of components like e.g.\ imaging artifacts or cell residues.

In order to evaluate whether the established random forest was overfitted to the training data set, a 10-fold cross validation was carried out.
Throughout all 10\,folds, results were very similar, resulting in low standard deviations. 
This indicates the RF model to be independent from the underlying training data and suitable for the classification of unknown and new data.

Furthermore, the prediction quality was studied separately for all five materials.
Since various materials affect platelets differently, shape and size of adherent platelets varies from material to material.
Within the manually classified ground truth data set, components from all material groups were included.
Nevertheless, we needed to rule out that our proposed method predicts platelets on some materials better than on others.
In general, ACCs, MCCs and FPRs are in a similar range for all five materials. 
Some outliers are present in the MCC for the area of components with values down to 0.51 (PCU\textsubscript{cast}).
This drop of MCC can be explained by the extremely low fraction ($<\SI{1}{\percent}$) of the positive class for this material and is not due to a RF weakness.
The only material-related effects are observed for FNRs.
Whereas differences between materials are small for the number of components, the area of components shows strong differences for some materials.
PDMS and PET show FNRs in the range of the total FNR, contrarily, the other three materials result in considerably lower FNRs.
On the one hand, this is in line with the optical impression that PDMS presents the highest number of difficult to classify positive components (e.g.\ proteins).
On the other hand, the results are in contrast to the general finding that FNRs increase with decreasing portions of positive components.
In conclusion, the proposed method predicts platelets on different materials well with the limitation that an elevated number of uncharacteristic positive components results in higher FNRs.
This can be overcome in the future by either improving the quality of images and thus reducing the appearance of such components or 
by including more of these specific components into the ground truth of the random forest training data set.

A general source of error is the supervised setup, which might be biased by the manual classification of a single person.
The multi-person comparison shows that the RF developed within this study is not noticeably biased in this regard.
Within the different person-person and person-RF combinations there were no differences either for the number or the area of components in terms of ACC, MCC and error rates.
Of course, some combinations showed slightly better results than others, but there was no clear pattern for any combination being superior to all others.
Error rates are generally lower compared to the final RF values, this is probably due to the smaller data subset with only about 1\,000\,components and consequently less variability within this subset.

Finally, the whole hemocompatibility study was analyzed by means of the newly developed automatized analysis method, which combines image segmentation and a RF.
The results are generally in line with the expectations, since materials with a known outcome were tested \citep{Braune2017}.
PTFE shows the most platelet adhesion and activation as the platelet covered area is significantly higher compared to all other materials \citep{Braune2017, Chandy2000, Freeman2018}.
PDMS shows a very low platelet adhesion affinity, which corresponds to its reported good short-term hemocompatibility \citep{Spiller2007, Khorasani2004}.
The remaining materials are situated in between this range as expected \citep{Chandy2000, Clauser2014}. 
In contrast to a previous study, no significant differences were observed for PET compared to PDMS \citep{Braune2017}.
This might be due to the new analysis technique that allows for analyzing the whole material sample area.
Without automation of platelet counting and characterization, only small fractions of the material sample can be analyzed, which might lead to varying results.

Overall, all statistics and results give evidence that the established RF in combination with the segmentation algorithm is a suitable tool for the automated classification of platelets on fluorescence microscopy images.
However, a RF is always tailored to a specific problem due to the underlying ground truth data used for training.
Therefore, the transferability to other microscopy techniques (e.g.\ Scanning Electron Microscopy, Laserscanning Microscopy), staining methods (e.g.\ specific P-selectin staining)
and materials has to be evaluated before the RF can be applied to those images.
Only if the appearance of platelets is the same in terms of size, stained area, resolution, etc., the RF can be applied without the need for a new training based on the new ground truth.
This restriction applies also for the segmentation algorithm.
If, for example, the image quality, resolution or platelet appearance differs severely to the here-in used images, segmentation might become imprecise.
In this case, segmentation parameters have to be adjusted to the present case.

A further improvement of the developed automated platelet analysis would overcome the manual image inspection prior to the image segmentation.
This could be achieved by an additional pre-processing step prior to platelet classification.
As an example, a convolutional neural network could be used for the first image sorting since it is capable for large data-set classification \citep{Castilla2018, Pang2017}.
Furthermore, a more distinct platelet characterization with regard to the activation state of the platelets would reveal more details of the material-blood interaction.

Although there is room for improvements, the developed analysis method is the first approach for an automatized platelet analysis for hemocompatibility evaluation.
Therewith, it offers the possibility for fast and reproducible in-vitro experiment analyses in the future.
This is a first and important step towards the standardization of biomaterial hemocompatibility evaluations, 
which will contribute to further progress in the development of blood-contacting medical devices.

\section{Conclusion}
\label{sec5}
In this study, we developed a new tool for the analysis of platelet adhesion on biomaterials after in-vitro testing.
The new method includes automatized image segmentation as well as platelet characterization by means of a Mumford-Shah model and a Random Forest, respectively.
Therewith, the limitations of so far used (semi-)manual platelet count techniques are overcome.
Based on this automatized analysis method, future hemocompatibility investigations will become more comparable and reproducible 
and thus allow for standardized and generally applicable hemocompatibility assessment.

\section*{Appendix}
Supplementary material is available online.

\section*{Acknowledgment}
The authors like to thank Dr. Doris Keller from the University Medical Center RWTH Aachen University for the blood withdrawal.
Calculations were performed with computing resources granted by RWTH Aachen University under project rwth0314.
The authors have no competing interests to declare. 

This study was partly funded by the INTERREG Program V-A Euregio Maas-Rhine of the European Union (Grant Number 2016/98602).

B. Berkels was funded in part by the Excellence Initiative of the German Federal and State Governments through grant GSC 111.

\bibliographystyle{elsarticle-num-names}
\bibliography{Literatur_PUB_ThrombanalysisAutomation}

\end{document}